\documentclass
[showpacs,superscriptaddress,prl,twocolumn,tightenlines,balancelastpage,10pt,a4paper]{revtex4}
\usepackage{amsmath}
\usepackage{amsfonts}
\usepackage{amssymb}
\usepackage{graphicx}%

\begin{document}

\title{Entanglement assisted spin-wave atom interferometer}
\author{Yu-Ao Chen}
\affiliation{Physikalisches Institut, Ruprecht-Karls-Universit\"{a}t
Heidelberg, Philosophenweg 12, 69120 Heidelberg, Germany}
\affiliation{Hefei National Laboratory for Physical Sciences at Microscale and Department of Modern Physics, University of Science and Technology of China, Hefei, Anhui 230026, China}
\author{Xiao-Hui Bao}
\affiliation{Physikalisches Institut, Ruprecht-Karls-Universit\"{a}t
Heidelberg, Philosophenweg 12, 69120 Heidelberg, Germany}
\affiliation{Hefei National Laboratory for Physical Sciences at Microscale and Department of Modern Physics, University of Science and Technology of China, Hefei, Anhui 230026, China}
\author{Zhen-Sheng Yuan}
\affiliation{Physikalisches Institut, Ruprecht-Karls-Universit\"{a}t
Heidelberg, Philosophenweg 12, 69120 Heidelberg, Germany}
\affiliation{Hefei National Laboratory for Physical Sciences at Microscale and Department of Modern Physics, University of Science and Technology of China, Hefei, Anhui 230026, China}
\author{Shuai Chen}
\affiliation{Physikalisches Institut, Ruprecht-Karls-Universit\"{a}t
Heidelberg, Philosophenweg 12, 69120 Heidelberg, Germany}
\author{Bo Zhao}
\affiliation{Physikalisches Institut, Ruprecht-Karls-Universit\"{a}t
Heidelberg, Philosophenweg 12, 69120 Heidelberg, Germany}
\author{Jian-Wei Pan}
\affiliation{Physikalisches Institut, Ruprecht-Karls-Universit\"{a}t
Heidelberg, Philosophenweg 12, 69120 Heidelberg, Germany}
\affiliation{Hefei National Laboratory for Physical Sciences at Microscale and Department of Modern Physics, University of Science and Technology of China, Hefei, Anhui 230026, China}
\date{\today}
%%%%%%%%%%%%%%%%%%%%%%%%%%%%%%%%%%%%%%%%%%%%%%%%%%%%%%%%

\begin{abstract}
We report the observation of phase-super resolution in a
motion-sensitive spin-wave (SW) atom interferometer utilizing a
NOON-type entanged
state. The SW interferometer is implemented by generating
a superposition of two SWs and observing the interference between
them, where the interference fringe is sensitive to the atomic
collective motion. By heralded generation of a second order
NOON state in the SW interferometer, we clearly observe the interference
pattern with phase super-resolution. The demonstrated SW interferometer
can in principle be scaled up to highly entangled quantum state, and thus
is of fundamental importance to implement quantum-enhanced-measurement.
\end{abstract}


\maketitle

The optical interferometer  \cite{Dowling08} and the atom interferometer  \cite{Cronin08}
have become
essential tools for measuring position, displacement or
acceleration. In these devices, a classical light
pulse or an ensemble of neutral atoms are coherently split and
recombined in space domain or time domain by applying mechanical
or optical gratings. The gravity or platform rotation will cause a
motion-sensitive phase shift, which can be measured from the
interference fringes.

As is well known, by exploiting suitable quantum entanglement,
e.g. NOON state or GHZ state, the measurement precision can be
improved \cite{Giovannetti04,Dowling08}. For optical
interferometers, the principle of the quantum-enhanced-measurement
has been demonstrated by exploiting photonic NOON state, where
phase super-resolution \cite{Walther04,Mitchell04} and phase
super-sensitivity \cite{Nagata07} have been observed for N=4.
However, generating atomic NOON state or GHZ state and exploiting
them in atom interferometers are still challenging for current
technology. Although entanglement up to eight atoms has been
achieved in ion traps \cite{Haffner05}, this system can not be
exploited in conventional motion-sensitive atom interferometers
due to the fact that the atoms have to be ionized and confined in
a small volume.

Here we propose and demonstrate a motion-sensitive
entanglement-assisted SW atom interferometer. In our protocol,
instead of using superposition of single atoms, we exploit
superposition of the collective excited states, i.e. the SW, to
implement the interferometer. The collective motion of the atoms
will cause a phase shift, which can be measured by converting the
SW into photons. The SW interferometer is experimentally
demonstrated by using cold atomic ensembles, where the collective
motion is introduced by a radiation pressure force. Further, we
heralded generate a second order NOON state in the SW
interferometer. The motion sensitivity is proved to be twice
higher, which clearly shows the phase super-resolution. Our method
can lead to phase super-sensitivity with the improvement of
coherence time of the SWs. Moreover, the SW interferometer can in principle
be scaled up to highly entangled states, which might be useful in
inertial sensing.

\begin{figure}[ptb]
\begin{center}
\includegraphics[width=3in]{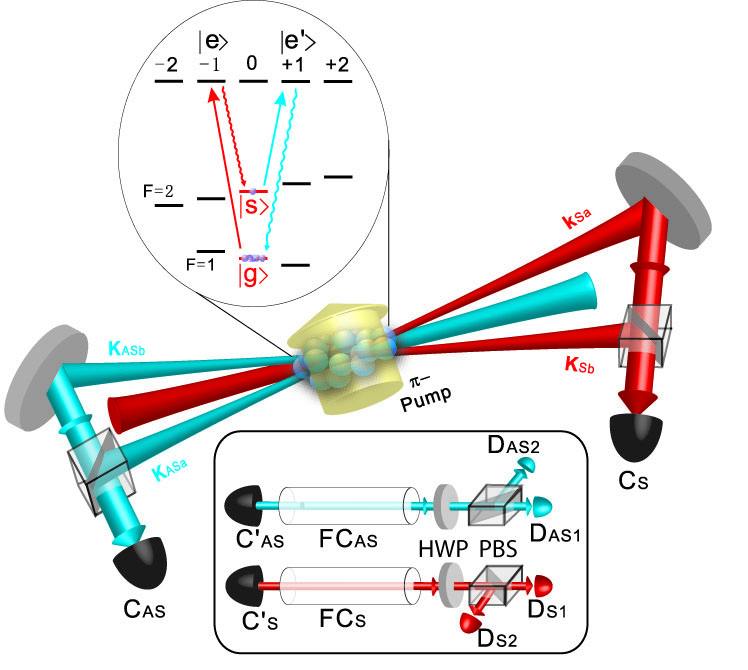}
\end{center}\caption{ Experimental setup of the SW
interferometer. The inset at the top shows the relevant Zeeman
levels for the $ \vert 5 S_{1/2}\rangle\rightarrow\vert 5 P_{1/2},
F^{\prime}= 2\rangle$ transition of $^{87}$Rb atoms. Before the
experimental cycles we optically pump the atoms in $\vert g
\rangle$ by applying two pumping lights  (See Appendix for detail.), one of which
is shining in from the lateral side ($\pi$-Pump) to introduce
radiation pressure force. A weak $\sigma^{-}$ polarized write
pulse is applied to generate two modes of SW and Stokes fields via
spontaneous Raman transition $|g\rangle \rightarrow |e\rangle
\rightarrow|s\rangle$. The Stokes fields are collected at the
angle of $\pm 0.6^{0}$ relative to the write beam, combined on a
PBS and directed into a single mode fiber via a fiber coupler
(C$_S$). After a controllable delay, a strong $\sigma^{+}$
polarized read beam induces the transition $|s\rangle
\rightarrow|e^{\prime}\rangle \rightarrow |g\rangle$, converting
the two SWs into two anti-Stokes fields, which are overlapped in
another PBS and then directed to a coupler (C$_{AS}$). Passing
through two filter cells (FC) respectively, the anti-Stokes photon
from C$^{\prime}_{AS}$ and Stokes photon from C$^{\prime}_S$ are
then sent to the polarization analyzers combined with half wave
plate (HWP), polarized-beam splitter (PBS) and single photon
detectors (D), as illustrated in the inset at bottom. Filter cells
are properly pumped in order to absorb the remaining leakage from
read or write beams while to be transparent for the signals.}\label{setup}
\end{figure}

To illustrate the working scheme of SW interferometer, we consider
a cold atomic cloud with the $\Lambda$-type level structure shown
in the Fig. 1. All the atoms are initially optically pumped to
$\vert g\rangle$. An off-resonant $\sigma^{-} $ polarized write
pulse coupling the transition $|g\rangle\rightarrow|e\rangle$ with
wave vector $\mathbf{k}_{\mbox{\tiny W}}$ is applied to the atomic
ensemble along the axial direction, inducing spontaneous Raman
scattering. Two Stokes fields with $\sigma^{-}$ polarization and
wave vector $\mathbf{k}_{\mbox{\tiny S}a}$ and
$\mathbf{k}_{\mbox{\tiny S}b}$ are collected at an angle of
$\theta_{a,b}=\pm\theta$ relative to the write beam. The
atom-light field in each mode can be expressed as \cite{DLCZ}
\begin{equation}
|\Psi \rangle_i \sim
|0\rangle_{i}^{\mbox{\tiny S}}|0\rangle_i+\sqrt{\chi_i}
|1\rangle_{i}^{\mbox{\tiny S}}|1\rangle_i+\chi_i
|2\rangle_{i}^{\mbox{\tiny S}}|2\rangle_i+
O(\chi_i^{3/2}),
\end{equation}
where $\chi_a=\chi_b\ll1$ is the excitation probability of one
collective spin excitation in mode $i$ ($i=a, b$), $\vert j
\rangle_{i}^{\mbox{\tiny S}}$ denotes the the Stokes field S$_i$
with photon number $j$, while $\vert j\rangle_i=S^{\dagger
j}_i\vert 0\rangle_i$ denotes the $j$-fold collective spin
excitation in mode $i$, with $\vert 0\rangle_i=\bigotimes_l\vert
g\rangle_l$ the vacuum,
$S^{\dagger}_i=\frac{1}{\sqrt{M_i}}\sum_{l}e^{i
\Delta\mathbf{k}_i\cdot \mathbf{ r }^i_l}\vert s\rangle_l\langle
g\vert$ the creation operator of SW$_i$, where $\Delta
\mathbf{k}_i=\mathbf{k}_{\mbox{\tiny W}}-\mathbf{k}_{\mbox{\tiny
S}i}\simeq \mathbf{k}_{\mbox{\tiny W}}\sin\theta_i$ is the wave
vector of SW$_i$ and $\mathbf{r}^i_l$ denotes the coordinate of
the $l$-th atom in mode $i$.

The two Stokes fields are rotated to be horizontally ($\vert
H\rangle$) and vertically ($\vert V\rangle$) polarized for mode
$a$ and $b$ respectively, and are combined on a polarized beam
splitter (PBS). The half-wave plate (HWP$_{s}$) is set to
22.5$^{\circ}$ to measure the Stokes photons under $\vert
\pm\rangle=\frac{1}{\sqrt 2}(\vert H\rangle \pm\vert V\rangle)$
basis. Neglecting high order excitations, a click on detector
D$_{\mbox{\tiny {S1}}}$ or D$_{\mbox{\tiny {S2}}}$ will project
the atomic ensembles into the superposition state
\begin{equation}
\vert\Psi\rangle=\frac{1}{\sqrt 2}(\vert
1\rangle_a\vert0\rangle_b\pm \vert 0\rangle_a\vert1\rangle_b).
\label{1st}
\end{equation}
Such a SW superposition state can be exploited to implement the
Mach-Zehnder interferometer.

Assume the atoms undergo a collective motion, e.g. motion caused
by gravitational acceleration, described as $\mathbf{r}^{\prime
i}_l=\mathbf{r}^{i}_l+\mathbf{r}_c$. Since $\vert
1\rangle_i=S^{\dagger}_i\vert
0\rangle_i=\frac{1}{\sqrt{M_i}}\sum_{l}e^{i\Delta \mathbf{k}\cdot
\mathbf{ r }^i_l}|g...s_l...g\rangle$, after the collective motion
the SW will change to $\vert 1 ^{\prime}\rangle_i= e^{i
\phi_i}\vert1\rangle_i$ with $\phi_i= \Delta\mathbf{k}_i\cdot
\mathbf{ r }_c$, where
$\Delta\mathbf{k}_a=-\Delta\mathbf{k}_b\equiv\Delta\mathbf{k}$.
Thereby, we obtain
\begin{eqnarray}
\vert\Psi^{\prime}\rangle &=& \frac{1}{\sqrt 2}(e^{i\Delta \phi}\vert 1\rangle_a \vert0\rangle_b\pm e^{-i \Delta\phi}\vert 0\rangle_a\vert1\rangle_b) \nonumber\\&\sim&\frac{1}{\sqrt 2}(\vert 1\rangle_a \vert0\rangle_b\pm e^{-i~ 2 \Delta\phi}\vert 0\rangle_a\vert1\rangle_b),
\label{evolve}
\end{eqnarray}
with $\Delta\phi=\Delta\mathbf{k}\cdot \mathbf{ r }_c$. It can be
readily seen that the collective motion of the atoms is mapped to
a relative phase in the superposition state. This phase and thus
the collective motion can be measured by converting the SW back
into photons and observing the interference pattern of the SW
interferometer. In this way, if $\Delta \mathbf k$ is set in the
direction of the gravity, one can measure the gravitational
acceleration. Moreover, any quantity regarding to center of mass
displacements can be measured. The measurement precision is
corresponding to the sensitivity of the interferometer determined
by length of the wave vector $\Delta \mathbf k$, which is
controllable in practice.

\begin{figure}[ptb]
\begin{center}
\includegraphics[width=3.5in]{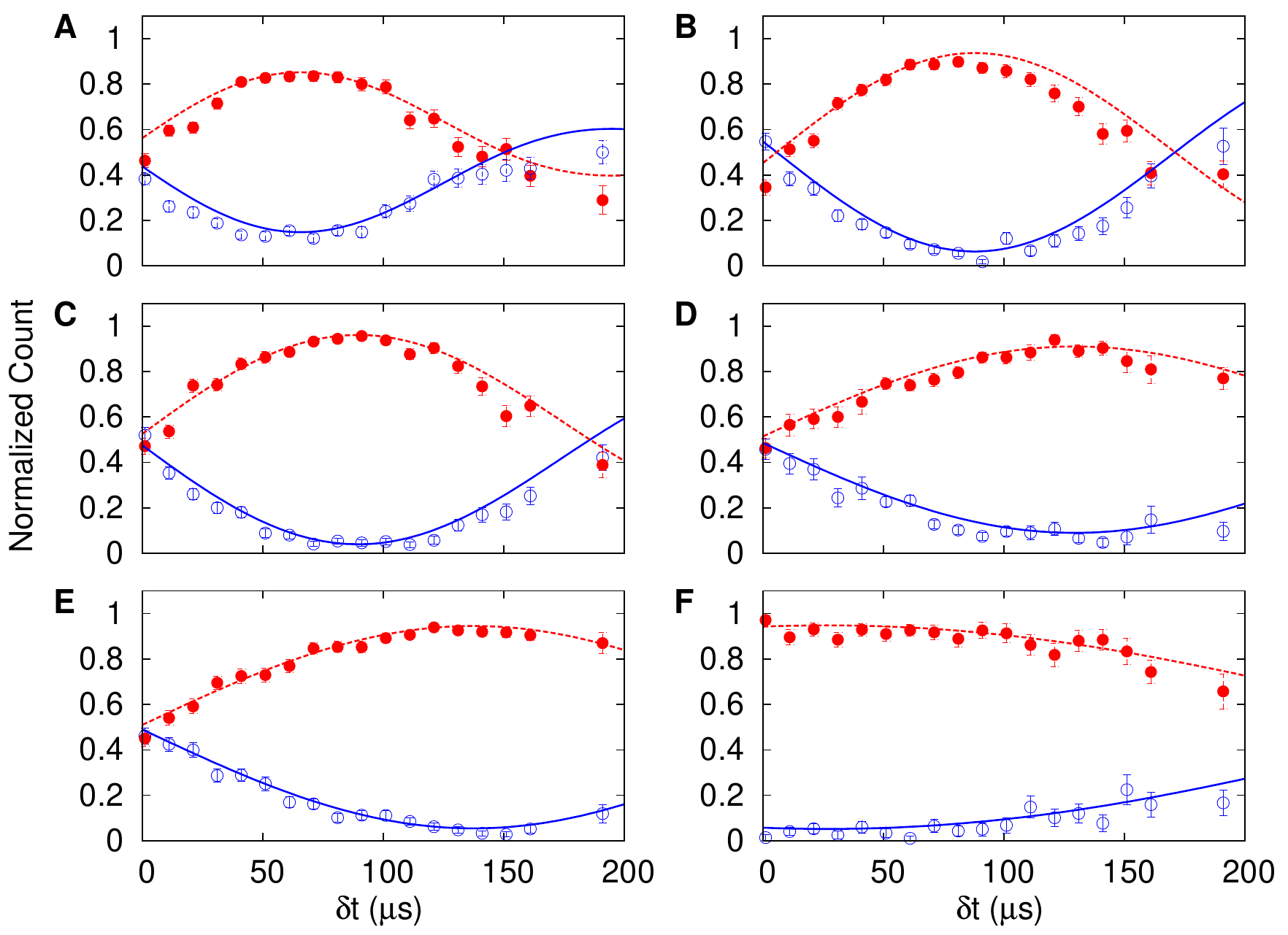}
\end{center}\caption{The fidelity of anti-Stokes field on $\vert
+\rangle$ as a function of $\delta t$. Solid circles (open
circles) represents the measured fidelity of anti-Stokes field on
$\vert +\rangle$ on condition of a click of Stokes field on state
$\vert + \rangle$ ($\vert -\rangle$). The power of the $\pi$-Pump
is: \textbf{(A)} 6 mW, \textbf{(B)} 4.5 mW, \textbf{(C)} 3mW, \textbf{(D)}
1.5 mW, \textbf{(E)} 0.75 mW, \textbf{(F)} 0 mW. The experimental data
are jointly fitted by using $f_{+|+}(\delta t)=(0.5+a\sin^2(\pi
\frac{t}{T}+\phi_0)e^{-\delta t^2/\tau^2})/(1+a\, e^{-\delta
t^2/\tau^2})$ and $f_{+|-}(\delta t)=(0.5+a\cos^2(\pi
\frac{t}{T}+\phi_0)e^{-\delta t^2/\tau^2})/(1+a\,e^{-\delta
t^2/\tau^2})$  (See Appendix for detail.). The evolution
period $T=\frac{\pi}{\Delta\mathbf{k}\cdot \mathbf{v_c}}$ is
measured to be \textbf{(A)} 317$\pm$18 $\mu$s, \textbf{(B)} 330$\pm$15
$\mu$s, \textbf{(C)} 378$\pm$14 $\mu$s, \textbf{(D)} 555$\pm$47
$\mu$s, \textbf{(E)} 591$\pm$30 $\mu$s, \textbf{(F)} 1177 $\pm$152
$\mu$s. Error bars represent statistical errors, which are $\pm1$
s.d.}\label{Phase}
\end{figure}

Such a single-excitation SW interferometer can be looked upon as
the first order of a NOON state \cite{Kok02}. Higher order NOON
state can be deterministically generated by using the linear
optical methods (See Appendix for detail.), described as
$$ \vert \mbox{NOON}
\rangle= \frac{1}{\sqrt 2}(\vert N\rangle_a\vert0\rangle_b+\vert 0
\rangle_a\vert N\rangle_b),$$
where $\vert N \rangle_i=S^{\dagger
N}_i\vert 0\rangle_i$ denotes the $N$-fold excitation in mode $i$
($i=a,b$). Thus the collective motion of the atoms will induce a
phase as
$$%\begin{equation}
\vert \mbox{NOON}^{\prime} \rangle\sim\frac{1}{\sqrt 2}(
\vert N\rangle_a\vert0\rangle_b +e^{-i~2N\Delta\phi}\vert 0\rangle_a\vert N\rangle_b).$$
Note that, although the method for
preparing NOON state is in principle extendable \cite{Walther04} to
arbitrary N, the efficiency of generating the desired NOON state
drops off exponentially \cite{Kok02} with N. In order to be more
efficient in employing entanglement, one can exploit multiple
atomic ensembles to prepare $N$-quanta Greenberger-Horne-Zeilinger
(GHZ) state \cite{GHZ89}
\begin{eqnarray}
\vert \mbox{GHZ}\rangle &=&\frac{1}{\sqrt 2}(\vert 1\rangle_{a_1}\vert0\rangle_{b_1}...\vert1\rangle_{a_{\tiny N}}\vert0\rangle_{b_{\tiny N}}\nonumber\\
&&+\vert 0\rangle_{a_1}\vert1\rangle_{b_1}...\vert0\rangle_{a_{\tiny N}}\vert1\rangle_{b_{\tiny N}}),\nonumber
\end{eqnarray}
which can be deterministically
generated in a scalable way \cite{Browne05,Barrett08}, thanks to
the built in quantum memory. The collective motion of the atoms
will induce a phase as
\begin{eqnarray}
\vert \mbox{GHZ}^{\prime}\rangle &\sim&\frac{1}{\sqrt 2}(\vert 1\rangle_{a_1}\vert0\rangle_{b_1}...\vert1\rangle_{a_{\tiny N}}\vert0\rangle_{b_{\tiny N}}\nonumber\\
&&+e^{-i~2N\Delta\phi}\vert 0\rangle_{a_1}\vert1\rangle_{b_1}...\vert0\rangle_{a_{\tiny N}}\vert1\rangle_{b_{\tiny N}}).\nonumber
\end{eqnarray}
Therefore, the SW
interferometer would be $N$ times more sensitive to the collective
motion with the help of these highly entangled states and thus
can be exploited to demonstrate the phase super-resolution
and phase super-sensitivity.

Demonstration of the SW interferometers critically depends on the
coherence time of the SW excitation. In the experiment, we
implement the SW interferometer with $^{87}$Rb atoms trapped in a MOT
at a temperature of about 100 $\mu$K. By exploiting the clock transitions
of, $|g\rangle=|5S_{1/2}, F=1,m_{F}=0\rangle$
and $|s\rangle=|5S_{1/2}, F=2,m_{F}=0\rangle$ as the two ground states to avoid the
deleterious effects induced by magnetic field, e.g. Lamor
procession or inhomogeneous broadening, we achieve the coherence
time of the SW of about 200 $\mu$s, which is limited by the
dephasing of the SW induced by atomic random motion \cite{Bo08}.
With such a coherence time, it is now possible to experimentally
study the motion sensitivity of the SW interferometer.

To show the motion sensitivity, we introduce a collective motion
during the pumping stage, where the atoms absorb photons from the
$\pi$-Pump light and the $2\rightarrow 2$ pump, and then decay
spontaneously. Since the $2\rightarrow 2$ pump is shined with the
cooler light from six directions, and spontaneous emission is in
arbitrary directions, on average they give no contribution to the
collective motion. While the $\pi$-Pump light from the lateral
side acts as a pushing laser, which causes a radiation pressure
force and accelerates the atoms \cite{Wohlleben01} until they are
pumped to $|g\rangle$. We denote the velocity acquired in this
process by $\mathbf{v_p}=v_p\hat{\mathbf e}_p$. Besides, the
unbalance of other lasers, i.e. cooler, repumper, and etc., will
also induce an initial velocity $\mathbf{v_0}$ when the atoms are
released. Note that, both of the modes $a$ and $b$ are accelerated
by the pushing laser. Therefore when the the superposition state
$\vert \Psi\rangle$ (Eq. \ref{1st}) is generated, it will evolve
to $\vert \Psi^{\prime}\rangle$ (Eq. \ref{evolve}) after a free
evolution time of $\delta t $, where $\Delta\phi=\Delta\phi(\delta
t)=\Delta\mathbf{k}\cdot\mathbf{v_{c}} \delta t$ with
$\mathbf{v_{c}=v_p+v_{0}}$ and in which the atomic random motion
is neglected.

\begin{figure}[ptb]
\begin{center}
\includegraphics[width=3in]{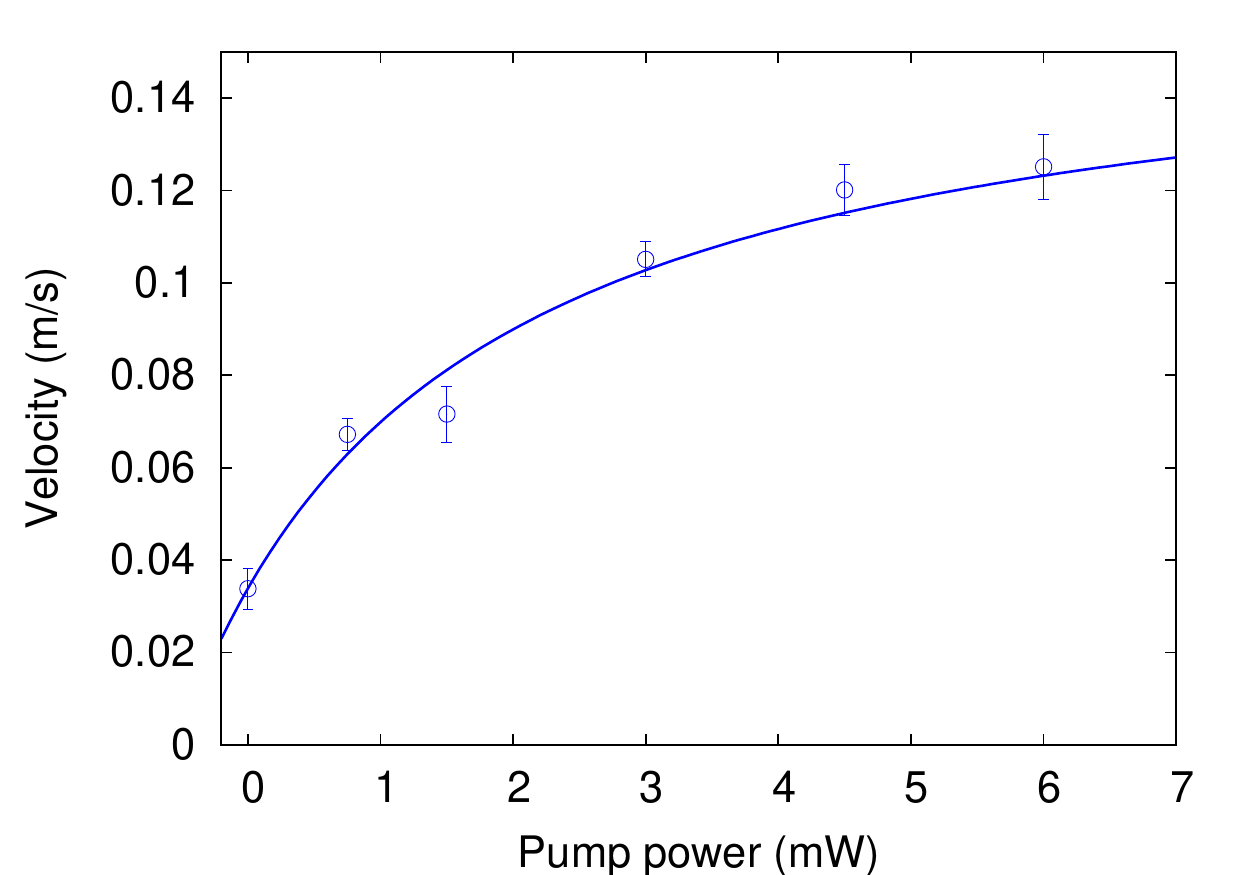}
\end{center}\caption{The atomic velocity as a function of the
$\pi$-Pump power. The initial velocity induced by the unbalance
of other lasers is about 0.03 m/s. The velocity acquired in the
pumping stage increases with the pumping power until reaching a
plateau of about 0.12 m/s. Error bars represent statistical
errors, which are $\pm1$ s.d.}\label{2nd}
\end{figure}

To measure $\Delta\phi(\delta t)$, a strong $\sigma^{+}$ polarized
read light, coupling the transition $|e^{\prime}\rangle
\rightarrow \vert g\rangle$, counter-propagating with the write
light, converts the collective excitations into $\sigma^{+}$
polarized anti-Stokes fields. The anti-Stokes fields from two
atomic ensembles are rotated to be perpendicular to each other and
combined on a PBS (Fig. 1), which can be described by
$\vert\Psi\rangle_{\mbox{\tiny {AS}}}\sim\frac{1}{\sqrt 2}(\vert H
\rangle_{\mbox{\tiny {AS}}}\pm e^{-i(2\Delta \phi(\delta t))} e^{i
(\phi_1+\phi_2)}\vert V\rangle_{\mbox{\tiny {AS}}}), $ where
$\phi_1$ ($\phi_2$) represents the propagating phase difference
between two Stokes (anti-Stokes) fields before overlapping. Note
that in the experiment, the total phase $\phi_1+\phi_2$ is
actively stabilized and set to a fixed value \cite{Yuao08}. The
interference pattern is observed by setting the HWP$_{AS}$ at
22.5$^{\circ}$ to detect the anti-Stokes fields under $+/-$ basis.
We change the power of the $\pi$-pump and measure the fidelity of
anti-Stokes field on $\vert +\rangle$ as a function of $\delta t$,
on condition of a click of Stokes field on state $\vert + \rangle$
(solid circles) and $\vert -\rangle$ (open circles). The
experiment results are shown in Fig. 2, where the data are fitted
by taking into account the retrieve efficiency and the background
noise (See Appendix for detail.). The collective motion can be obtained from the
period of the interference pattern
$T=\frac{\pi}{\Delta\mathbf{k}\cdot \mathbf{v_c}}$, which varies
from 300 $\mu$s to 1200 $\mu$s.

The velocity that the atoms acquired as a function of the
$\pi$-Pump power is shown in Fig. 3. One can see that the average
velocity will first increase with the $\pi$-Pump power, and reach
a plateau when the $\pi$-Pump is sufficient strong. This might be
related with the pumping efficiency in the pumping stage, since
when all the atoms are pumped to the $|g\rangle$, the atoms will
not absorb photons from $\pi$-Pump any more. Note that, the
population in other Zeeman sub-levels arising from insufficient
pump will not affect the interference pattern since the
decoherence time in other states is very shot (about microseconds)
due to inhomogeneous broadening.

\begin{figure}[ptb]
\begin{center}
\includegraphics[width=3in]{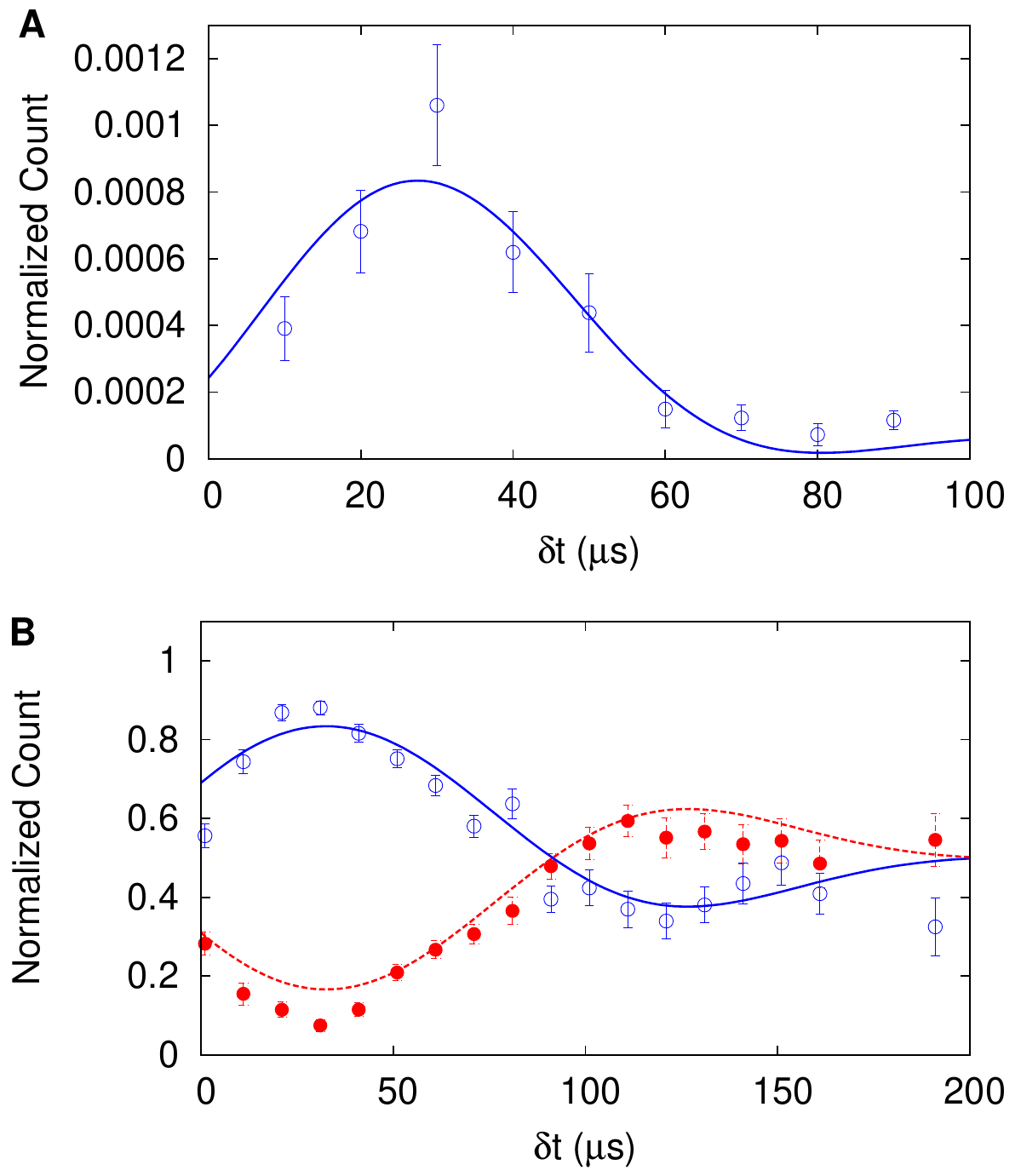}
\end{center}\caption{ Comparison of the performance of the first and
second order NOON state. \textbf{(A)}, The normalized coincidence
count for the second order NOON state as a function of $\delta t$.
The experimental data are fitted by using $c(\delta t)=b\sin
^{2}\left( \pi \frac{\delta t}{T^{\prime }}+\phi _{0}^{\prime
}\right) e^{-2\delta t^{2}/\tau ^{2}}+de^{-\delta t^{2}/\tau
^{2}}$  (See Appendix for detail.). The fitting evolution
period is $T^{\prime}=(94\pm12)$ $\mu$s. \textbf{(B)}, The
interference pattern for $N=1$ under the same condition. The lines
are the fitted with the same method as in Fig. 2. The fitting
evolution period is $T=(220\pm17)$ $\mu$s, which is about twice
high as $T^{\prime}$. Error bars represent statistical errors,
which are $\pm1$ s.d.}\label{2nd}
\end{figure}

To experimentally show the advantage of quantum entanglement, we
generate the second order NOON state, which can be described as
$\vert\Psi\rangle_{NOON}=\frac{1}{\sqrt 2}\left(\vert 2\rangle_a
\vert 0\rangle_b-\vert 0\rangle_a\vert2\rangle_b\right).$ This is
achieved by registering the coincidence count between single
photon detectors D$_{\mbox{\tiny {S1}}}$ and D$_{\mbox{\tiny
{S2}}}$ (See Appendix for detail.). Note that, such a state is deterministically
generated with the help of a feedback circuit \cite{Shuai06}. After
a free evolution time of $\delta t$, we have
$\vert\Psi\rangle_{NOON^{\prime}}\sim\frac{1}{\sqrt 2}(\vert
2\rangle_a \vert 0\rangle_b-e^{-i(4\Delta \phi(\delta t)}\vert
0\rangle_a \vert2\rangle_b).$ By converting the second order state
to anti-Stokes fields and measuring the coincidence count between
detectors D$_{\mbox{\tiny {AS1}}}$ and D$_{\mbox{\tiny {AS2}}}$,
we obtain the phase $\Delta \phi (\delta t)$. The normalized
coincidence count is shown in Fig. 4A. For comparison, we give the
interference fringe for the first order NOON state under the same
condition as shown in Fig. 4B, which is taken directly after the
measurement of the second order interference pattern. It can be
seen that the evolution period of second order NOON state (Fig.
4A, $T^{\prime}=(94\pm12)$ $\mu$s) is about 1/2 of the first order
(Fig. 4B, $T=(220\pm17)$ $\mu$s), which clearly shows the phase
super-resolution. Note that, these data are taken under $\pi$-pump
power of 6 mW. There's a slightly change of the initial velocity
of the atomic ensemble compared to the original condition, which
makes the first order evolving period slightly different to the
one given in Fig. 2A.

With emphasis, we note that NOON state is not only more sensitive
to the collective motion, but also more sensitive to
decoherence \cite{Dowling08,Huelga97}. Therefore, to achieve phase
super-sensitivity, the coherence time of our NOON state has to be
much larger than its free evolution
time \cite{Auzinsh04,Leibfried04}. However, in our experiment,
since the coherence time is comparable to the free evolution time,
the gain obtained by using N=2 NOON state is partially offset by
the corresponding faster decoherence, and thus we failed to
achieve phase super-sensitivity. It is expected that, our NOON
state will show the desired phase super-sensitivity with the
improvement of the coherence time.

In summary, we have proposed and demonstrated an entanglement
assisted SW atom interferometer. In the experiment, phase
super-resolution is clearly observed by exploiting the second
order NOON state. Higher order SW NOON state with small N can be
generated in current setup to further demonstrate the principle of
quantum-enhanced-measurement. Besides, since the quantum memory is
automatically built in our system, N-quanta SW GHZ state can be
deterministically generated in a scalable way, which is a distinct
advantage compared with photonic entanglement. The SW
interferometer might be used as an inertial sensor by
significantly improving several quantities, such as coherence time
and retrieve efficiency of the quantum memory, and the detection
efficiency. The coherence time of the quantum memory can be
substantially extended by, e.g., further reducing the temperature
or trapping the atoms in optical lattice \cite{Greiner02}. The
retrieval efficiency and the detection efficiency can be
respectively improved by exploiting an optical
cavity \cite{SimonJ07} and the photon-number-resolving
detectors \cite{James02,Imamoglu02}.

We acknowledge J. Schmiedmayer for useful discussions. This work was supported by the Alexander von Humboldt Foundation, the European Commission through the ERC Grant and the STREP project HIP, the National Fundamental Research Program (Grant No.2006CB921900), the CAS, and the NNSFC.

%%%%%%%%%%%%%%%%%%%%%%%%%%%%%%%%%%%%%%%%%%%%%%%%%%%%%%%%

\section{Appendix}

\renewcommand{\thefigure}{A\arabic{figure}}
 \setcounter{figure}{0}
\renewcommand{\theequation}{A.\arabic{equation}}
 \setcounter{equation}{0}

%%%%%%%%%%%%%%%%%%%%%%%%%%%%%%%%%%%%%%%%%%%%%%%%%%%%%%%%

\textbf{Measurement-induced NOON state.}
Our measurement-induced NOON state in spin-wave (SW) interferometer method
closely follows the scheme proposed by Nielsen and M{\o }lmer \cite%
{Nielsen07}, where the NOON state can be conditionally generated from two
pulsed type II optical parametric oscillators with only linear optical
devices. As shown in Supplementary Figure \ref{figS1}, two atom-light fields
(mode $a$ and $b$) can be written as $\vert\Psi\rangle_a\otimes\vert\Psi%
\rangle_b$, where
\begin{equation*}
|\Psi \rangle_i \sim\sum_N\sqrt{\chi_i}^N\vert N\rangle_{i}^ {%
\mbox{\tiny
S}}\vert N\rangle_i,
\end{equation*}
where $\chi_i\ll1$ is the excitation probability of one collective spin in
SW $i~(i=a,b)$, $\vert N \rangle_{i}^{\mbox{\tiny S}}$ denotes the the
Stokes field S$_i$ with photon number $N$ and $\vert N\rangle_i$ denotes the
$N$-fold collective spin excitation in SW $i$. The two light fields from
spontaneous Raman process selecting perpendicular polarization are guided on
a polarized beam-splitter (PBS). By adjusting the two modes $a$ and $b$ to
be equally excited ($\chi_a=\chi_b=\chi$), the atom-light fields can be
expressed as
\begin{eqnarray}
\vert\Psi\rangle_a\otimes\vert\Psi\rangle_b \sim \left(\sum_N\sqrt{\chi}%
^N\vert V\rangle^{\otimes N}\vert N\rangle_a \right)+\nonumber\\
\left(\sum_N\sqrt{\chi}%
^N\vert H\rangle^{\otimes N} \vert N\rangle_b\right)  \label{light-atom}
\end{eqnarray}

From the Eq. \ref{light-atom}, by a projection measurement of the combined
light field onto
\begin{equation}
\vert H\rangle^{\otimes N}-\vert V \rangle^{\otimes N},  \label{Stokes}
\end{equation}
with $H,V$ representing the horizontal and vertical polarization, the atom
field (the SW interferometer) is projected onto the NOON state \cite{A.Sun06},
\begin{equation}
\vert NOON\rangle\sim\vert N\rangle_a\vert 0\rangle_b-\vert 0 \rangle_a\vert
N\rangle_b.  \label{NOON}
\end{equation}
To do so, we rewrite Eq. \ref{Stokes} as,
\begin{equation}
\vert H\rangle^{\otimes N}-\vert V\rangle^{\otimes N}= \mathop%
\otimes\limits_{n=1}^{N}(\vert H\rangle-\vert V\rangle e^{i~2n\pi/N}).
\label{measure}
\end{equation}
From Eq. \ref{measure}, by properly setting the measurement device as shown
in Supplementary Figure \ref{figS1}, the coincidence between detectors D$_1$%
, D$_2$, ..., D$_N$ will project the SW interferometer onto the NOON state $%
\vert NOON\rangle$ (Eq. \ref{NOON}). Note that, although our
method for preparing NOON state is in principle
extendable \cite{A.Walther04} to arbitrary N, the efficiency of
generating the desired NOON state drops off
exponentially \cite{A.Kok02} as a function of N.

In the experiment, to generate the second order NOON state in SW
interferometer $\vert 2\rangle_a\vert 0\rangle_b-\vert 0\rangle_a\vert
2\rangle_b$, we need to perform the projection measurement of the Stokes
field onto $\vert H\rangle\otimes\vert H\rangle- \vert V\rangle\otimes\vert
V\rangle=(\vert H\rangle+\vert V\rangle) \otimes(\vert H\rangle-\vert
V\rangle)$. This can be simply achieved by setting the HWP$_{%
\mbox{\tiny{S}}}$ in Fig. 1 at 22.5$^0$ and measuring
the coincidence between detectors D$_{S1}$ and D$_{S2}$. This would be the
same as the complex measurement devices in Fig. %
\ref{figS1} when $N=2$. Note that, in our experiment, the
excitation probability $\chi\ll1$, and thus the contribution from
higher order excitations can be safely neglected.

%%%%%%%%%%%%%%%%%%%%%%%%%%%%%%%%%%%%%%%%%%%%%%%%%%%%%%%%

\begin{figure}[ptb]
\begin{center}
\includegraphics[width=3in]{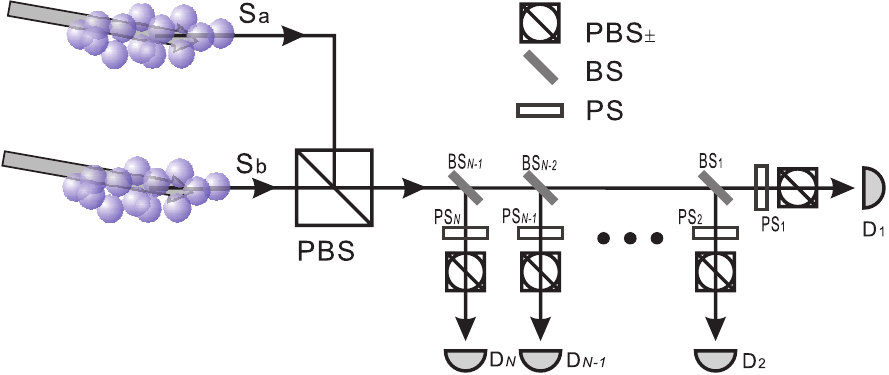}
\end{center}
\caption{
Measurement-induced NOON state preparation in SW interferometer. BS$%
_{i}$ denotes the beam-splitter with reflectivity of $\frac{1}{i+1}$. The
phase shifters PS$_j$ transform $\vert V\rangle$ to $-e^{i~2j \protect\pi%
/N}\vert V\rangle$. PBS$_{\pm}$ are polarizing beam-splitters oriented at 45$%
^0$, which reflects photons with $\vert -\rangle$ polarization and transmits
photons with $\vert+\rangle$. The coincidence between detectors D$_1$, D$_2$%
, ..., D$_N$ will project the SW interferometer onto the NOON state $\vert
NOON\rangle$ (Eq. \protect\ref{NOON}).}
\label{figS1}
\end{figure}

%%%%%%%%%%%%%%%%%%%%%%%%%%%%%%%%%%%%%%%%%%%%%%%%%%%%%%%%

\textbf{Experimental cycles.}
As shown in Fig. 1, in the experiment, the MOT is
loaded for 20 ms at a repetition rate of 40 Hz. The trapping magnetic field
and repumping beams are then quickly switched off. After 0.5 ms, the bias
magnetic field of about 3 G is switched on, whereas the cooler beams is switched off and
two pumping beams are switched on for 0.5 ms to optically pump the atoms to
the desired state $|g\rangle$. As the two pumping beams, one couples all the
magnetic transitions of $|5S_{1/2},F=2\rangle \rightarrow
|5P_{3/2},F^{\prime}=2\rangle$ (not shown in the figure), shining with
cooler light; the other pump ($\pi$-pump) couples the $\pi$ transition $%
|5S_{1/2},F=1\rangle\rightarrow|5P_{1/2},F^{\prime }=1\rangle$, shining in
from the side. Then, within another 4 ms, experimental trials (each
consisting of pumping, write and read pulses) are repeated with a
controllable period depending on the desired retrieval time. In order to
re-prepare the atoms to the desired state $|g\rangle$, we switch on the two
pumping beams in each experimental trial before write and read process, $%
|5S_{1/2},F=2\rangle \rightarrow |5P_{3/2},F^{\prime}=2\rangle$ pump for 1.2
$\mu s$ and $\pi$-pump for 0.9 $\mu$s.

To deterministically generate the second order NOON state, in each
experimental run, we keep shining the write pulses. Following each write,
the two pumping beams (clean pulses) are switched on to re-prepare the atoms
to the initial state $|g\rangle$. A click of coincidence between detectors D$%
_{S1}$ and D$_{S2}$ in Fig. 1 will stop the remaining
write pulses and clean pulses and the NOON state is induced in the SW
interferometer. After the SW state is converted back to anti-Stokes photons,
another experimental run restarts.

%%%%%%%%%%%%%%%%%%%%%%%%%%%%%%%%%%%%%%%%%%%%%%%%%%%%%%%%

\textbf{Fitting the experimental results.}
For the first order NOON state, considering the retrieval efficiency and the
background noise, the probability of detecting one photon in detectors D$_{%
\mbox{{\tiny AS1 }}}$ and D$_{\mbox{{\tiny AS2 }}}$, conditional on
detecting a Stokes field on state $|+\rangle $ are described by
\begin{eqnarray}
p_{\mbox{\tiny{AS1}}} &=&\gamma (\delta t)\cos ^{2}\left( \Delta \phi
(\delta t)+\phi _{0}\right) +\gamma _{b},  \notag \\
p_{\mbox{\tiny{AS2}}} &=&\gamma (\delta t)\sin ^{2}\left( \Delta \phi
(\delta t)+\phi _{0}\right) +\gamma _{b},  \notag
\end{eqnarray}%
where $\gamma (\delta t)=\gamma _{0}\exp (-\delta t^{2}/\tau ^{2})$ is the
overall retrieve efficiency, $\phi _{0}=-\frac{\phi _{1}+\phi _{2}}{2}$ and $%
\gamma _{b}$ denotes the background. The lifetime $\tau $ for single SW
excitation is measured from the decay from the correlation function as in
Ref. \cite{A.Bo08}. The fidelity of anti-Stokes field on $|+\rangle $ are
calculated to be $f_{+|+}(\delta t)=(0.5+a\cos ^{2}(\pi \frac{\delta t}{T}%
+\phi _{0})e^{-\delta t^{2}/\tau ^{2}})/(1+a\,e^{-\delta t^{2}/\tau ^{2}})$,
with $a=\frac{\gamma _{0}}{2\gamma _{b}}$ and $T=\frac{\pi }{\Delta \mathbf{k%
}\cdot \mathbf{v_{c}}}$ the evolution period. Similarly, the fidelity of
anti-Stokes field on $|+\rangle $ on condition of a Stoke photon on state $%
|-\rangle $ can be written as $f_{+|-}(\delta t)=(0.5+a\sin ^{2}(\pi \frac{t%
}{T}+\phi _{0})e^{-\delta t^{2}/\tau ^{2}})/(1+a\,e^{-\delta t^{2}/\tau
^{2}})$. The curves in Fig. 2 are joint fitting with $f_{+|+}$ and $f_{+|-}$%
, where $a$, $T$ and $\phi _{0}$ are fitting parameters.

For the second order NOON state, the coincidence count between D$_{%
\mbox{{\tiny AS1 }}}$ and D$_{\mbox{{\tiny AS2 }}}$ are measured, which
can be described as
\begin{equation}
p_{\mbox{\tiny{AS1}}}=\frac{1}{2}\gamma ^{2}(\delta t)\sin ^{2}\left(
2\Delta \phi (\delta t)+\phi _{0}^{\prime }\right) +2\gamma _{b}\gamma
(\delta t)+\gamma _{b}^{2},  \notag
\end{equation}%
where the first term is the two photon coincidence from the second order
excitation, the second term is noise coming from the coincidence between the
excitation and the background, and the third term is the coincidence between
background noise. The normalized data are the coincidence counts divided by
the sweeps, and are fitted by using
\begin{equation*}
c(\delta t)=b\sin ^{2}\left( \pi \frac{\delta t}{T^{\prime }}+\phi
_{0}^{\prime }\right) e^{-2\delta t^{2}/\tau ^{2}}+de^{-\delta t^{2}/\tau
^{2}},
\end{equation*}%
where we neglect the coincidence from background noise for
simplicity, $\tau $ is the
lifetime of the spin wave obtained from the decay from correlation, $b$, $%
d,\phi _{0}^{\prime }$ and $T^{\prime }=\frac{\pi }{2\Delta
\mathbf{k}\cdot \mathbf{v_{c}}}$ are fitting parameters. Note
that, from the first order data, we know that the background noise
is much smaller than the signal retrieved from the atom ensemble.
Thereby for the second order NOON state, the background
coincidence count can be safely neglected since it is the second
order small term.  This does not affect the conclusion of our
paper, since the lack of super-sensitivity in our experiment is
mainly caused by short coherence time of the SW.

\end{document}